\documentstyle[epsfig,11pt,psfig]{l-aa}

\begin{document}

\thesaurus{(08.02.1; 08.14.2; 13.25.5; 08.09.2)}

\title{BeppoSAX observations of the long period polar system 
V1309\,Ori \thanks{Also based on observations collected at the 
European Southern Observatory, La Silla, Chile}}

\author{D.~de Martino\inst{1} \and R. Barcaroli\inst{2} \and G. Matt\inst{2}
 \and M. Mouchet\inst{3,4} \and T. Belloni\inst{5} \and K. Beuermann\inst{6}
\and L. Chiappetti\inst{7} \and C. Done\inst{8} \and B.T. G\"ansicke\inst{6} 
\and F. La Franca\inst{2} \and  K. Mukai\inst{9}}

\offprints{D.~de Martino}

\institute{Osservatorio Astronomico di Capodimonte, Via Moiariello 16, 
          I-80131 Napoli, Italy
\and Dipartimento di Fisica, Universita' degli studi ``Roma Tre'', Via 
      della Vasca Navale 84, I-00146 Roma, Italy
\and DAEC, Observatoire de Paris, Section de Meudon, F-92195 Meudon Cedex,
          France
\and University Denis Diderot, Place Jussieu, F-75005 Paris, France
\and Astronomical Institute ``Anton Pannekoek'', University of Amsterdam, 
     Kruislaan 403, NL-1098 SJ Amsterdam, The Netherlands
\and Universit\"ats Sternwarte, Geismarlandstr. 11, D-37083 G\"ottingen, 
     Germany
\and Istituto di Fisica Cosmica CNR, Via Bassini 15, Milano, Italy
\and Department of Physics, University of Durham, South Road, Durham 
DH1 3LE, UK
\and NASA/GSFC, Code 668 Greenbelt MD 20771, USA }

\date{Received Oct. 22, 1997; accepted Jan. 6, 1998}

\maketitle

\begin{abstract}

We present BeppoSAX observations of the peculiar long period polar 
system V1309\,Ori (RXJ0515.6+0105). The source was detected
simultaneously at soft and, for the first time, at hard X-rays 
with the LECS and the MECS detectors. Both, the LECS and the MECS 
light curves are irregular with a bursting/flaring type behaviour
indicating inhomogeneous accretion onto the white dwarf. This peculiar 
variability, together with an extreme high soft-to-hard X-ray luminosity
ratio,  indicates that in V1309\,Ori accretion occurs predominantly in
highly compressed chunks or ``blobs'' of matter.  From coordinated 
ESO optical spectroscopy, we find indications that the magnetic field
strength of the white dwarf is $<$ 70\,MG, not expected either
from the 8\,hr orbital period synchronism or from the strong soft-to-hard
X-ray ratio suggesting alternative solutions for sustaining synchronism
in this system.

\keywords{Stars: binaries: close 
       -- Stars: cataclysmic variables 
       -- X-rays: stars
       -- Stars: individual: V1309\,Ori = RXJ0515.6+0105}
\end{abstract}

\section{Introduction}

Polars or AM Her stars are Cataclysmic Variables containing a
synchronously rotating magnetic ($\rm B \sim$ 10-230 \,MG) 
white dwarf accreting from a late-type
secondary star. The strong magnetic field of the accreting white
dwarf dominates the accretion flow  and channels it  towards the
magnetic polar regions where a strong stand--off shock is
produced.  The hot post-shock plasma emits hard X-rays, partially
absorbed and re-emitted from the surface at soft X-ray,  EUV and UV
wavelengths, as well as cyclotron emission which is observed at optical and
IR wavelengths (Cropper 1990; Beuermann 1997). An independent soft X-ray
component can be produced by the infall of dense plasma packets or blobs 
which penetrate deep into the atmosphere of the white dwarf and heat the 
photospere from below to a few $10^5$ K (Kuijpers \& Pringle 1982). 

Polars have orbital periods in the range from 80\,min to se\-ve\-ral
hours, V1309\,Ori being the longest period system with P$_{\rm
orb}=$ 7.98\,hr. (Garnavich et al. 1994, Walter et al. 1995, Shafter
et al. 1995, Buckley \& Shafter 1995). V1309\,Ori is an
eclipsing polar showing a deep and variable primary minimum and a
shallower secondary one (Garnavich et al. 1994, Shafter et al. 1995,
Buckley \& Shafter 1995).  A magnetic field of $\leq$ 60MG was
inferred from cyclotron features in the optical and IR and from
optical polarimetry (Garnavich et al. 1994, Shafter et al. 1995,
Buckley \& Shafter 1995, Harrop-Allin et al 1997), although a higher
magnetic field was expected for such a long period synchronous system.

The soft X-ray emission, as observed by ROSAT, is strongly variable on
timescales down to few seconds in a bursting like activity, interpreted by 
Walter et al. (1995) as evidence for ``blobby'' accretion.
The ROSAT data also indicated an X-ray variability at the $\sim$
8\,hr orbital period. Separate ROSAT pointings revealed that
V1309\,Ori shows, as most other polars do, a long-term variability in its
X-ray flux due to changes in the mass transfer rate. The soft
X-ray emission was found to be consistent with a black-body emission
at 50\,eV.  No constraints on the hard component could be
established from the ROSAT data.  The extraordinary long orbital
period of V1309\,Ori makes it a key object
to test theories for synchronization of the white dwarf while its
pronounced short-term variability at X-ray wavelengths makes it a test
case for the theory of ``blobby'' accretion.

In the framework of a program aiming to detect simultaneously the soft
and hard X-ray emission in polars with the recent X-ray facility given
by the BeppoSAX satellite, we present new X-ray observations of
V1309\,Ori obtained during the BeppoSAX AO1-Core Program together
with coordinated optical spectroscopy collected at ESO.

\section{\it BeppoSAX observations}

The BeppoSAX X-ray satellite (Boella et al. 1997a) observed 
V1309\,Ori on October 5 1996 with the 
set of the four co-aligned Narrow Fields Instruments (NFI), co\-ve\-ring 
the wide energy range 0.1 - 300\,keV, used 
as prime pointing instruments. The source has been detected only with 
the Low Energy Concentrator Spectrometer (LECS), a gas scintillator spectrometer
covering the 0.1-10 keV energy range (see Parmar et al. 1997 for
detailed description) and the three units of the Medium Energy
Concentrator Spectrometers covering the 1.3-10 keV range (see Boella
et al. 1997b).  The source was not detected by the two high energies 
instruments, the HPGSPC (High Pressure Gas Scintillation 
Proportional Counter) and the PDS (Phoswich Detection System).
Due to LECS instrumental problems and operational limitations 
during the satellite 
orbit, the resulting effective on--source exposure was 10ksec, whilst 
MECS observation, carried out without significant problems, resulted in an
effective on--source exposure of $\sim$60ksec. In Table 1 we report
the observation log.

\begin{table}     
\caption[]{Journal of observations.}
\begin{flushleft}
\begin{tabular}{lll}
\hline
\noalign{\smallskip}
Instrument & Date - UT$_{\rm start}$ & Date - UT$_{\rm end}$  \\
    &        ~~~~~~~~~~ hh:mm &      ~~~~~~~~~~ hh:mm \\
\noalign{\smallskip}
\hline
\noalign{\smallskip}
LECS & 1996 Oct.5 15:24 & Oct.6 09:46 \\
\noalign{\smallskip}
MECS & 1996 Oct.5 15:24 & Oct.6 22:15  \\
\noalign{\smallskip}
\smallskip

ESO DFOSC & 1996 Oct.7 08:58 & \,~~~~~~~~09:18  \\ 
\noalign{\smallskip}
ESO DFOSC & 1996 Oct.8 04:56 & \,~~~~~~~~05:21 \\ 
 & &   \\
\hline
\end{tabular}
\end{flushleft}
\end{table}

LECS and MECS observations have been pre-processed at the BeppoSAX 
Software Data Center (SDC). The three MECS unit event files have been
summed after equalization to unit 1. Hereafter the name  MECS 
will refer to  the sum of the three units. 
LECS and MECS spectra and light curves have been extracted 
from circular regions  with radii of 4' and 2' around the
centroid source ima\-ge respectively.  
Since the background in both instruments was found to be stable 
along the satellite orbit, it has been measured from blank sky pointings  from 
the same regions of source count extraction. 
The average background count rates are
$5.62 \pm 0.19 \cdot 10^{-3}$ cts\,s$^{-1}$ and $ 2.71 \pm 0.08 \cdot
10^{-3}$ cts\,s$^{-1}$ for LECS and MECS, respectively.  Light curves
and spectra have been extracted using the XSELECT package, while the XSPEC and 
XRONOS packages
have been used to perform spectral and temporal analysis.

\begin{figure*}          
\mbox{\epsfxsize=18.cm\epsfbox{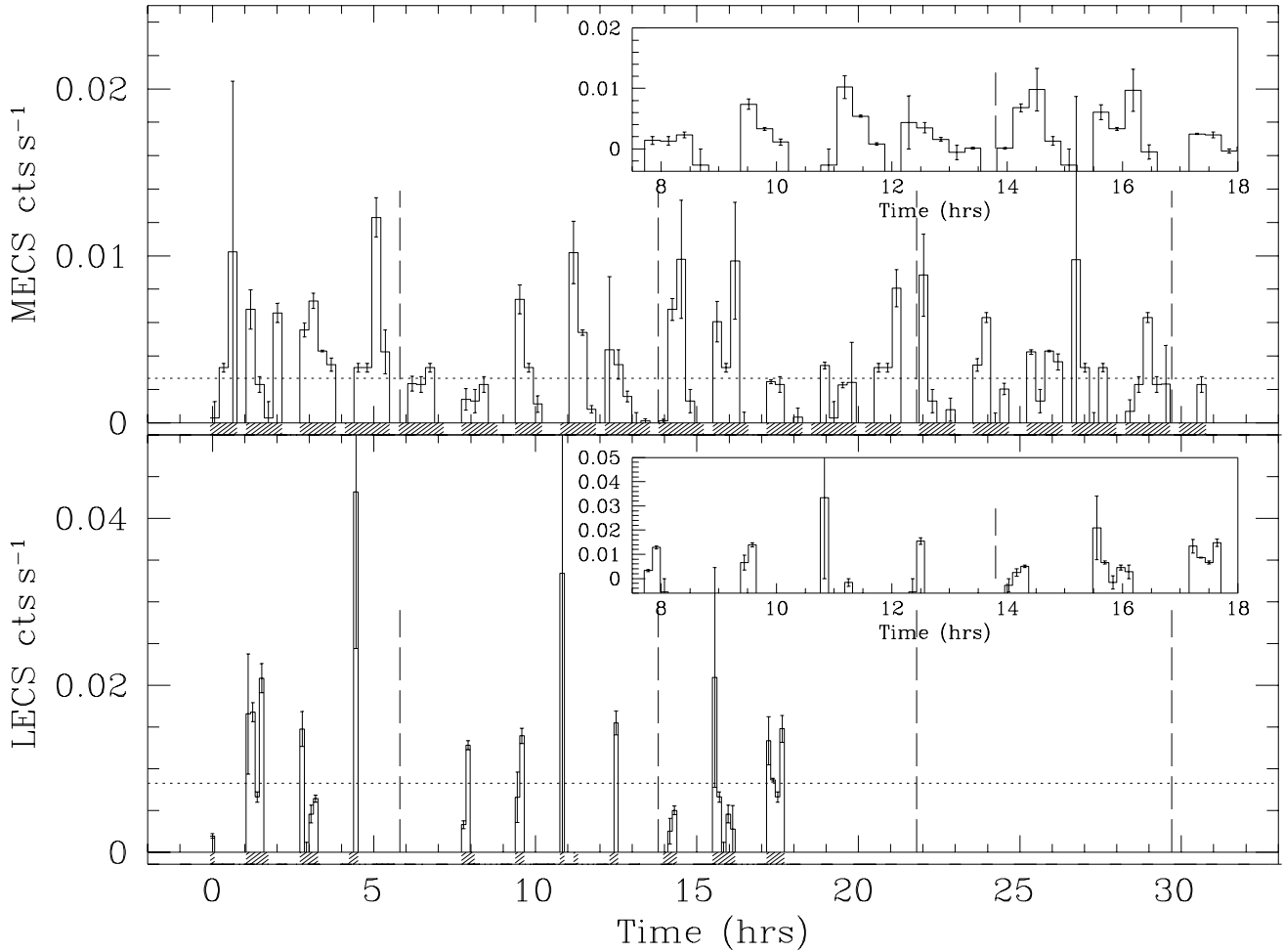}}
\caption[ ]{The net light curves in the LECS (lower panel) and MECS (upper 
panel) detectors averaged in 500\,sec and 1000\,sec bins respectively. 
The mean count rates are also shown with a dashed line. Enlargements of
are also reported in each panel where it is apparent
that the source occasionally switches off in both detectors. 
The 98\,min satellite orbital gaps as well as those related to LECS functioning
are marked as shaded areas. The times of mid-eclipse using the refined
ephemeris by Buckley \& Shafter (1995) are marked with a
dashed line. Note that counts below zero are due to the uniform background subtraction procedure.}
\end{figure*}

\subsection{The X-ray light curves}

LECS and MECS light curves are shown in Fig. 1 where the data have
been binned in 500\,s and 1000\,s bins respectively.  The average net
count rate in LECS is $7.42 \pm 1.25 \cdot 10^{-3} \rm cts\,s^{-1}$
and $2.70 \pm 0.35 \cdot 10^{-3} \rm cts\,s^{-1}$ in MECS. 
As it will be discussed in par.2.2, the LECS detector is dominated by 
the soft X-ray component of V1309\,Ori as also apparent from 
the hardness ratio [2-10\,keV] - [0.1-2\,keV] /[0.1-2\,keV] + 
[2-10\,keV] = -0.22. Thus MECS and LECS ligth curves are indicative
of the variability in the hard and soft X-ray emissions respectively. 
In both light curves the source shows a flaring/bursting activity, as  
it is apparent from the enlargements reported in each panel.  
This peculiar behaviour is composed by 
discrete events interleaved by faint states and occasional switching 
off of the source. The source is $\sim 55-60\%$ of the time at a low
level of emission. In these light curves the flaring activity has
typical timescales $\le 30$\,mins, consistent to that
observed during previous ROSAT observations  although the latter 
allowed the detection of variations on timescales down to 
few seconds (Walter et al. 1995). In these binned light curves the source
is observed to vary by a factor up to 6 and to 3 in the LECS and MECS
respectively with respect to the average level. 
Even though the poor coverage with the LECS detector prevents a statistical
correlation study between the temporal behaviour of the soft and hard
X-ray components, it is clear from Fig.\,1, that the bursts/flares do not
necessarily coincide in the two light curves. The low level of the source
in the instruments prevents any study of the behaviour in the hardness ratios
within each instrument energy bands so to ascribe a spectral trend to this
peculiar variability.  
It is worth mentioning that these BeppoSAX observations do not reveal 
the 8\,hr bursting periodicity as inferred from ROSAT data by Walter et 
al. (1995). Due to satellite orbital gaps and the low level of the source,
which hampers a finer time binning  of the data,  we are 
unable to detect the eclipse of the white dwarf (Fig.1).

\begin{figure}          
\mbox{\epsfysize=8.cm\epsfxsize=7.cm\epsfbox{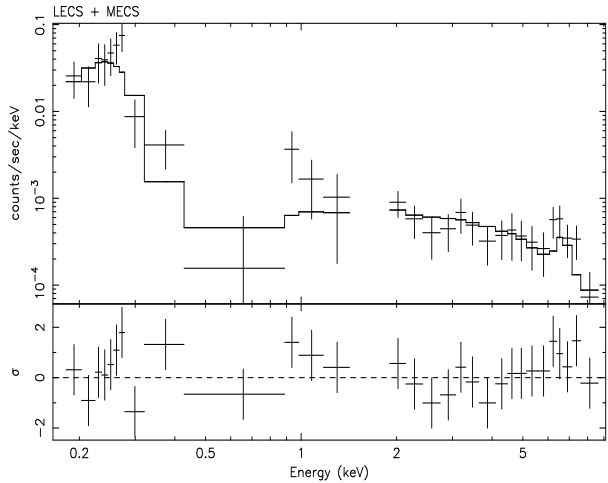}}
\caption[ ]{Average MECS and LECS spectra together with the composite
spectrum consisting of a  30\,eV black-body and a 10\,keV Raymond-Smith 
models. The excess of flux at 0.3\,keV is likely due to the LECS 
calibration matrices.}
\end{figure}

\subsection{The X-ray spectrum}

The soft X-ray component of V1309\,Ori was detected in the
ROSAT band (Walter et al. 1995) but no constrains on the hard X-ray 
component could be found. The extension to higher energies than those 
covered by ROSAT ($\geq$ 2\,keV) allows the detection of the hard 
X-ray emission  for the first time.
The average X-ray composite LECS and MECS spectrum of V1309\,Ori is shown
in Fig. 2 together with a two component best fit consisting of a black-body
kT$_{\rm \small \rm bb}$ \normalsize = 30$^{+30}_{-21}$ eV and a
Raymond-Smith (Raymond \& Smith 1977) kT$_{\rm \small \rm RS}$
\normalsize = 10$^{+\,\infty}_{-5}$ keV models, assuming solar abundances
(A=1) ($\chi^{2}$ = 20.6, Dof=25). The quoted errors are 90$\%$
confidence for one interesting parameter ($\Delta \chi^2$ = 2.7). 
The hydrogen
column density is found to be 4 $\cdot 10^{20} \rm cm^{-2}$, poorly  
constrained,   but compatible with the value of 5.4\,$\pm\, 1.3 \cdot
10^{20} \rm cm^{-2}$, found from ROSAT data (Walter et al. 1995). 
While ROSAT observations are consistent with a $\sim$ 50 \,eV black-body, we
still find a 30eV temperature fixing N$_{\small \rm H}$ \normalsize 
to 5.4\,$\cdot 10^{20} \rm cm^{-2}$. 
With the derived spectral fits parameters, the measured flux 
in the 0.1-2 keV range is 5.8$\cdot 10^{-13} \rm erg\, cm^{-2}\ s^{-1}$
while in the 2-10\,keV range it is 
2.1$\cdot  10^{-13} \rm erg\, cm^{-2} s^{-1}$.
The black-body bolometric flux 
ranges between $2-5 \cdot 10^{-11} \rm erg\, cm^{-2} s^{-1}$ for 
$\rm kT_{\rm bb} =30-50$\,eV and $\rm N_{H} = 4-5.4\cdot 10^{20} \rm cm^{-2}$.
On the other hand, from the emission measure of the optically 
thin component at 10\,keV, assuming a Gaunt factor of 1.2, we find a 
bolometric flux of 
$3.1 \pm 0.7 \cdot 10^{-13} \rm erg\,cm^{-2} s^{-1}$. This implies a very high soft-to-hard X-ray ratio
$\rm F_{bb} /F_{RS}$ ranging between 65 and 160. We also compared the
soft-to-hard flux ratio in V1309\,Ori with those observed in other polars 
(Beuermann \& Woelk 1996, Beuermann 1997) in the ROSAT band  0.1-2.4\,keV.
We fitted the BeppoSAX data adopting for the optically thin plasma a  
temperature  of 20\,keV, a 25\,eV black-body 
and $\rm N_{H}$ to 4$\cdot 10^{20} \rm cm^{-2}$, obtaining 
$\rm F_{bb} /F_{RS}= 430$, the highest ratio ever observed in any system.

We furthermore note that the hydrogen
column density as indicated by the ROSAT and our data 
is much lower than the total interstellar galactic column 
in the direction of the source, $\sim 10^{21} \rm cm^{-2}$, which would be
expected from the distance estimate of 1.5\,kpc (Harrop-Allin et al. 1997)
and it is more compatible with a $\sim$ 500\,pc distance as suggested by 
Buckley \& Shafter (1995). The distance $\rm d$ of V1309\,Ori
can be inferred from the R band ratio of the observed brightness of the 
secondary star at mid-eclipse (given in Shafter et al. 1995) and the surface 
brightness of an M0-1 star. As $\rm d \propto R_{2}$
and the radius of the Roche-lobe filling star depends on its mass as 
$\rm R_{2} \propto M^{1/3}_{2}$, we find 
$\rm d \simeq 830\, (M_{2}/M_{\odot})^{1/3}\,pc$. 
The  secondary star has driven out of thermal equilibrium 
(Frank et al. 1995) and while its temperature is that of an M0-1 star 
($\sim 3800\,\rm K$), its luminosity and mass are not easy to estimate. 
We then consider a possible range for $\rm M_{2}$
between 0.25 and 0.80\,$\rm M_{\odot}$, which gives $\rm d \simeq 525 - 777\,pc$.
 The distance
of 1.5\,kpc by Harrop-Allin (1997) then appears to be overestimated.
For an average distance of 650\,pc, the bolometric black-body luminosity 
is then $1-2.5 \cdot 10^{33} \rm erg\,s^{-1}$ and that of the 
thermal optically thin plasma is  1.6$\pm 0.4 \cdot 10^{31} \rm erg\, s^{-1}$.

\section{The Optical Observations}

Optical spectroscopy has been carried out on
October 7/8, at the Danish 1.54m ESO telescope equipped with
DFOSC+CCD/LORAL 2kx2k spectrograph configuration operated with grism 4
covering the range between 3550 and 9000 \AA\ with a nominal
dispersion of 220 \AA\ mm$^{-1}$. A slit of 2.5" has been used
resulting into a 16.9 \AA\ resolution, as measured from Ar-He arc
ca\-li\-bra\-tion lines.  Two spectra with exposures times of 20\,min and
25\,min have been acquired centred at phases 0.48 and 0.99
respectively, (using the refined eclipse ephemeris of Buckley $\&$
Shafter (1995)). Given the shape of the optical light curve (Garnavich et 
al. 1994), the spectrum at phase 0.48 falls on the egress phases of the
  secondary minimum.
Data have been reduced using the MIDAS package for standard spectroscopic
reduction procedures including bias, flat-fielding and wavelength
calibration. The latter has been shifted using the sky line at 5577.4
\AA\ to correct for deflections introduced by the telescope.  
The spectra have been extracted using the optimized method by Horne
(1986).  Flux calibration has been performed using observations of the
standard star LTT 1788.

\begin{figure}          
\mbox{\epsfxsize=8.8cm\epsfbox{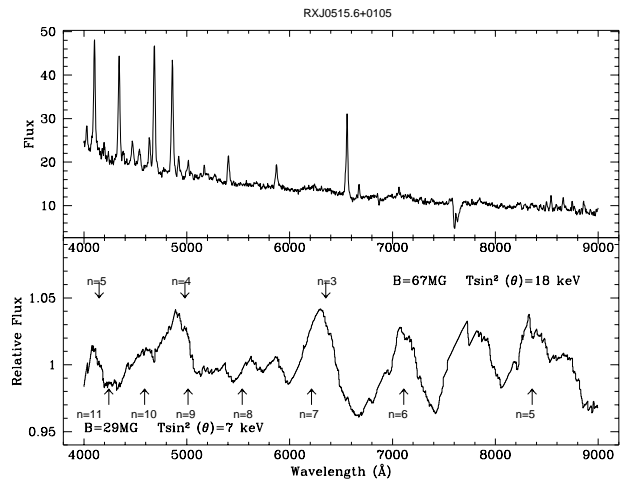}}
\caption[ ]{{\bf Upper panel:} The difference spectrum between 
spectra at 0.48 and 0.99 orbital phases. Ordinates are in units of
$10^{-16} \rm erg\,cm^{-2}\, s^{-1}\, \AA^{-1}$.
{\bf Lower panel:} The smoothed 
nor\-ma\-li\-zed cyclotron spectrum and the expected positions for the two
best solutions of B and $\rm T\,sin^{2}(\theta)$ as described in the text.}

\end{figure}

Both spectra are very similar in shape and intensity to those observed
at phases 0.21 and 0.01 by Shafter et al. (1995). While the
out-of-eclipse spectrum shows the strong Balmer, He\,II and He\,I
emission lines characteristic of an AM Her star, the eclipse spectrum 
displays the secondary star absorption spectrum typical of a M0 dwarf
(Garnavich et al. 1994, Buckley et al. 1995).  In order to analyze
the accretion induced spectrum we removed the latter component by
subtracting the eclipse spectrum. We note that the secondary star
absorption features, especially the TiO band at 7600 \AA, are not completely
removed. 
This difference spectrum, shown in the upper panel of
Fig. 3, displays a strong ratio of $\rm H_{\beta}/\rm
H_{\alpha}$=1.5 while $\rm H_{\beta}/\rm H_{\gamma} \sim$
1. Also, He\,II 4686\, \AA\, is similar in flux to that of H$_{\beta}$. In
order to detect cyclotron features we removed strong and faint
emission lines by means of gaussian fits while the residual secondary
star absorptions have been excised interpolating the adjacent
continuum.  Here we note that  the ellipsoidal modulation
might introduce a bias or a systematic error  in the 
cyclotron spectrum.
It is expected that the contribution of the secondary star 
at the two phases is different. This  can
be additionally affected by X-ray heating which is however difficult
to quantify. We estimate the difference in the optical flux of the
two minima due to ellipsoidal va\-ria\-tions neglecting at first 
approximation heating effects.
Following the work of Bochkarev et al. (1979) for a mass ratio of 1
(0.8$\leq$ q $\leq$ 1.4 (Shafter et al. 1995)), a gravity darkening
parameter $\beta$=0.25 (at optical wavelengths), a radius ratio of the
secondary star to that of the polar Roche lobe between 0.9-1, 
an inclination angle of 90$^{\rm o}$ and a limb darkening ranging between 0.6 to 1, 
the difference of the two minima ranges from 0.03\,mag
to a peak value of 0.16\,mag. For an average value of 0.09\,mag, this 
difference converts into an error of $\sim 2\%$ in the cyclotron flux. 
We then applied a median filtering and  normalized the spectrum 
using a composite polynomial function fitted to the observed
continuum. The resulting spectrum is shown in the lower panel of Fig. 3.
Five weak features are detected which might
be interpreted as cyclotron emissions.  
Although their detection is at the one sigma level,
(the noise in this spectrum $\sim 4\%$), our spectrum is 
very similar in intensity to those derived by 
Garnavich et al. (1994) and Shafter et al. (1995) at other phases. 
In order to obtain estimates on the magnetic field
strength and temperature $\rm T sin^2(\theta)$ where T is the temperature
in keV and $\theta$ is the viewing angle to the field lines
(Cropper et al. 1989), we 
fitted the observed positions, but  no consistent solution is found 
to explain all of them. Retaining the two strongest features at
$\sim$ 4842\AA ~and 6282\,\AA, we find 
B=67 $\pm$ 1\,MG and $\rm T sin^2(\theta)= 18
\pm$ 1\,keV (quoted errors are at 68$\%$ confidence).  
These features correspond to the 4$^{\rm th}$(4912 \AA)
and 3$^{\rm rd}$(6283 \AA) harmonics respectively. 
On the other hand, a solution with B= 29$\pm$ 1\,MG and
 $\rm T sin^2(\theta) = 7 \pm$1\,keV would account for the humps at 8357\,\AA, 
 7106\,\AA,  and 6282\,\AA (n=5,6 and 7 respectively). This would also  
predict harmonics up to n=11, whose pre\-se\-nce is rather difficult to assess 
(see Fig.\,3). 
We note that the the broad hump at 7762 \AA,
which is very close to the secondary star absorption, is likely due
to the removal procedure.  The first solution
matches with the results from optical spectroscopy and polarimetric 
observations of Buckley $\&$ Shafter (1995), Shafter et al.  (1995) and 
Harrop-Allin et al. (1997), while the second one is compatible with the 
low field estimate of 33\,MG by Garnavich et al. (1994) 
(though they find a very low temperature). Although we cannot be conclusive 
given the method we used and the single spectrum we base our analysis on, 
we conservatively prefer the first solution of a 67\,MG field white dwarf 
which accounts for the strongest humps. The lower field solution cannot 
be excluded, but, however,  it needs further observational basis, 
i.e. higher S/N spectroscopy possibly at different orbital phases.
 These data, differently from Shafter et al.'s (1995) statement, 
give indication that cyclotron emission is present around 
secondary minimum.

\section{Discussion and Conclusions}

During the BeppoSAX observations V1309\,Ori was at a flux level 
comparable to that observed during 
one of the ROSAT pointings in 1991 when the source was found to be highly 
variable. A bursting-on/off behaviour, which can be decomposed in 
quiet intervals interleaved with strong flares,  has been observed by 
BeppoSAX not only in the soft X-rays but also in the hard X-ray emission. 
Such variability indicates 
that the activity is due to occasional increases of accretion onto the white 
dwarf as also suggested by Walter et al. (1995). However the fact that 
the source sometimes switches off 
in both hard and soft X-rays is a strong indication of 
a highly  inhomogeneous accretion. Although fla\-ring acti\-vi\-ty has been
observed in other polars like BL Hyi (Beuermann \& Schwope 1989) 
and QS Tel (Rosen et al. 1996), V1309\,Ori is the first system to show
such a marked variability also in the hard X-rays.

Its X-ray spectrum consists of a soft and a hard component 
which can be described by a 30\,eV black-body and a 
10\,keV optically thin plasma emission (Raymond-Smith model). 
This first detection of the hard X-ray
component allows us to derive an extremely high soft-to-hard X-ray bolometric
flux ratio of $\sim$ 65-160.
Such a large soft X-ray excess indicates that in V1309\,Ori most
of the kinetic energy is emitted from a shock buried deep in
the white dwarf atmosphere, at optical depths greater than one even in the
hard X-rays. Radiation tran\-sfer then reprocesses the
primary thermal bremsstrahlung into soft X-rays emitted from the 
surface.  We estimate the mean accretion rate assuming
that the bulk of the accretion luminosity is irradiated in the soft 
X-rays: $\rm L_{bb}=1-2 \cdot10^{33} erg\, s^{-1} \sim 
G \dot M\, M_{wd} R_{wd}^{-1}$.
For a standard white dwarf of 0.6$M_{\odot}$ and a radius of 
8$\cdot10^{8} \rm cm$, 
$\rm \dot M=1.5-3.2\cdot 10^{-10} M_{\odot}\,yr^{-1}$ which is within
the observed range of other polars.
A buried shock may form if the white dwarf in V1309\,Ori possesses a
very high magnetic field $\geq$ 150\,MG and cyclotron cooling is so
efficient that the stand--off shock collapses.  Such a high field would
be indicated by the 8\,hr orbital synchronism for a 0.6 M$_{\odot}$ 
white dwarf (Patterson 1994), but it is not 
confirmed by the observations (see below).

Another possibility is that the matter is accreted predominantly at
high local mass flow rates in form of discrete blobs which
penetrate deep in to the atmosphere of the white dwarf (Kuijpers \&
Pringle 1982). Considering that there is no observational evidence for
an extremely high magnetic field and that the X-ray light curves
display a strong bursting/flaring character, we favour the idea of
``blobby'' accretion to explain the soft X-ray excess in V1309\,Ori.
However, these BeppoSAX observations indicate that the conventional  
``blobby '' accretion picture should be modified to account for 
blobs also to produce hard X-rays.

>From optical spectroscopy and
polarimetry a magnetic field strength up to 60\,MG was derived
(Garnavich et al. 1994, Buckley $\&$ Shafter 1995, Shafter et
al. 1995, Harrop-Allin et al. 1997). 
Our  optical spectrophotometry acquired during primary
minimum and close to the secondary mi\-ni\-mum, also suggests a
magnetic field of $<$70\,MG indicating a 
low field white dwarf, and thus 
suggesting different solutions to  the mechanism of maintaining 
synchronism. Indeed Frank et al.  (1995) proposed that
synchronism in V1309\,Ori, can be sustained if the secondary 
star possesses a relatively high field ($\geq$
1\,kG). On the other hand a light white dwarf $\leq 0.47 M_{\odot}$ is
required from the synchronism and for an upper limit of 70\,MG as also 
suggested by Shafter et al. (1995). Is then possible that two mechanisms are
occurring in V1309\,Ori: 
the action of a
higher field of the secondary star as proposed by Frank et al. (1995)
as well as a lighter accreting white dwarf. 

\medskip

In summary, while our new X-ray observations indicate that V1309\,Ori
represents an extreme case of ``blobby'' accretion, the optical
data support a relatively low field white dwarf thus making
this system a test case for further detailed theoretical work.

\begin{acknowledgements}

We acknowledge BeppoSAX SDC team for providing pre-processed event files
and for their constant support and advice in data reduction.
GM acknowledges financial support from ASI.

\end{acknowledgements}


\begin{thebibliography}{}


\bibitem{} Beuermann K. 1997, in proc. {\it Perspectives in High Energy 
Astronomy $\&$ Astrophysics}, International Coll. to Commemorate the Golden 
Jubilee of TIFR, in press.

\bibitem{} Beuermann K. \& Schwope A. 1989, A\&A 223, 179.

\bibitem{} Beuermann K. \&  Woelk U. 1996, in {\sl Cataclysmic Variables and
related objects\/}, IAU Coll 158, eds. A. Evans, \& J.H., Wood, p.~199
(Dordrecht: Kluwer)

\bibitem{}Boella G. et al. 1997a, A$\&$AS 122, 299. 


\bibitem{}Boella G. et al. 1997b, A$\&$AS 122, 327.


\bibitem{} Bochkarev N.G. et al. 1979, Sov. Astron. 23, 8. 


\bibitem{} Buckley D.A.H $\&$ Shafter A. 1995, MNRAS 275, L61.


\bibitem{}Cropper M. et al. 1989, MNRAS 236, 29p.



\bibitem{}Cropper M. 1990, Space Science Reviews 54, 195.


\bibitem{}Frank J. et al. 1995, ApJ 453, 446.

\bibitem{}Garnavich P.M.  et al.  1994, ApJ 435, L141.



\bibitem{} Harrop-Allin M.K. et al. 1997, MNRAS 288, 1033.


\bibitem{} Horne K. 1986, PASP 98, 609.

\bibitem{} Kuijpers  J. \& Pringle J.E. 1982, A\&A 114, L4

\bibitem{} Parmar A.N.  et al. 1997, A\&AS 122, 309.


\bibitem{} Patterson J. 1994, PASP 106, 209.

\bibitem{} Raymond J.C. \& Smith M.G. 1977, ApJSS 35, 419.

\bibitem{}Rosen S.R. et al. 1996, MNRAS 280, 1121.






\bibitem{}Shafter A.W. et al. 1995, ApJ 443, 319.



\bibitem{}Walter F.M. et al.  1995, ApJ 440, 834.



%
%
%
%
%
%
%
%
%
%
%
%
%
%
%




\end{thebibliography}
\end{document}